# Heat capacities of nanostructured wurtzite and rock salt ZnO: challenges of ZnO nano-phase diagram

Konstantin V. Kamenev,[1] Alexandre Courac,[2,3] Petr S. Sokolov,[4] Andrei N. Baranov,[5] Felix Yu. Sharikov,[6] and Vladimir L. Solozhenko [4,*]

[1] School of Engineering and Centre for Science at Extreme Conditions, The University of Edinburgh, EH9 3JZ Edinburgh, United Kingdom

[2] IMPMC, Sorbonne Université, CNRS, IRD, Muséum National d'Histoire Naturelle, 75005 Paris, France

[3] IUF, Institut Universitaire de France, 75005 Paris, France

[4] LSPM–CNRS, Université Sorbonne Paris Nord, 93430 Villetaneuse, France

[5] Department of Chemistry, Moscow State University, 119991 Moscow, Russia

[6] Saint Petersburg Mining University, 199106 Saint Petersburg, Russia

* Correspondence: vladimir.solozhenko@univ-paris13.fr

**Abstract:** Low-temperature heat capacities ($C_p$) of nanostructured rock salt (rs-ZnO) and wurtzite (w-ZnO) polymorphs of zinc oxide were measured in the 2–315 K temperature range. No significant influence of nanostructuring on $C_p$ of w-ZnO has been observed. The measured $C_p$ of rock salt ZnO is lower than that of wurtzite ZnO below 100 K and is higher above this temperature. Using available thermodynamic data, we established that the equilibrium pressure between nanocrystalline w-ZnO and rs-ZnO is close to 4.6 GPa at 300 K (half as much as the onset pressure of direct phase transformation) and slightly changes with temperature up to 1000 K.

## 1. Introduction

The low-temperature heat capacity measurements are required for understanding the phase transformations and construction of low-temperature regions of the equilibrium pressure – temperature ($p$-$T$) phase diagrams of elements and compounds, especially those important for mineralogy and explorative materials science (e.g. search for new materials under high pressure). At the same time there are a number of problems that arise during the heat capacity measurements for analysis of equilibria at low temperatures and high pressures, such as

(1) the direct (adiabatic) measurements at ambient pressure are time-consuming, costly and most importantly require a significant amount of pure compound (typically hundreds of μg), which makes explorative studies and study of phases recovered from very high pressures difficult (the pressure achievable in the laboratory is inversely proportional to the reaction volume);

(2) the results of typical Physical Property Measurement System (PPMS) measurements are limited in temperature (~350 K) and can be influenced by systematic errors due to either poor thermal contact between the sample and the heater-thermometry platform; and/or between the grains in a polycrystalline sample (porosity, bad sintering, reduced inter-grain contacts, etc.);

(3) systematic errors can also arise due to the water present in two forms – (i) bound into crystal structure during the sample synthesis and (ii) absorbed from air due to the porosity of the sample;

(4) low thermal stability of some high-pressure phases (such as rs-ZnO) can limit possible measurements to low temperatures, while the data of interest may lay at much higher temperatures, so the reliable extrapolation procedures are required; and finally,

(5) heat capacity data collection at high pressure remains rather exotic, very time-consuming, available in arbitrary units and accessible only in a very narrow $p$-$T$ domain.

Usually the construction of the phase diagrams at high enough temperatures (subject to the material and the pressure) is not considered be of an importance once the reversible transitions between phases are observed. However, in some cases even at high temperature the crystallization of a new allotrope/polymorph is not reversible and, moreover, may occur outside the domain of thermodynamic stability (e.g. in the case of boron [1] or boron oxide [2]). The fundamental question of low-temperature stability can be hardly resolved without reliable low-temperature heat capacity data. Thus, the convincing thermodynamic analysis of materials at extended $p$-$T$ regions of the phase diagram is hardly possible without low-temperature data on $C_p$ at ambient pressure and the methods of its extrapolation to high temperatures and high pressures.

Zinc oxide ZnO is a functional material that can be used both in industrial applications and for basic research in common w-ZnO form [3], as well as in high-pressure rs-ZnO form that is a material of interest for bright blue luminescence [4]. rs-ZnO can be recovered at ambient conditions as individual (nanostructured) phase, up to the volume of ~100 mm$^3$ per individual high-pressure synthesis experiment [4,5]. The recovery of such a substantial sample volume allowed to conduct the measurements of a wide range of physical properties - previously unavailable - such as, for example, the standard enthalpy of the w-ZnO-to-rs-ZnO phase transformation $\Delta H$(298.15 K) = 11.7±0.3 kJ mol$^{-1}$ [6]. At the same time, the reported data should be taken with precaution, since the bulk samples with micro-sized grains are available only for w-ZnO. Recent advances in the diverse synthetic routes of rs-ZnO materials [4] allowed us to obtain samples of chemical purity for reliable measurements of the heat capacities of both polymorphs for consistent side-by-side comparison.

Here we report low-temperature heat capacities for both nanostructured ZnO polymorphs, consistent with observable phase transformation between polymorphs at high temperature (when pressure hysteresis between direct and inverse transformations become negligible). We established that the nanostructured samples contain the absorbed water (problem (3) of the list above) which is the major source of measurement errors that can be quite significant and should be considered when physical property measurements are conducted. Accounting for the exact amount of absorbed water (by using thermal gravimetric analysis) allowed us to obtain the accurate estimate of heat capacities and to evaluate the equilibrium pressure between w-ZnO and rs-ZnO polymorphs at high pressure and room temperature. No significant impact of nanostructuring on heat capacity has been observed for w-ZnO.

## 2. Experimental Section

rs-ZnO can be recovered at ambient conditions only when nanopowders of w-ZnO are used as precursors [4]. Preparation of starting w-ZnO nanopowders by methods of solution chemistry [4] does not allow synthesis of ZnO nanoparticles free of surface chemical groups that can strongly impact the heat capacity measurements and subsequent free energy calculations. Milling techniques also does not allow obtaining the samples of high crystal perfection for measurements of sufficient thermodynamic quality. To avoid any organic or inorganic impurity - except small amount of absorbed water – we synthesized w-ZnO nanopowder by thermal decomposition of zinc peroxide $ZnO_2$ (Prolabo, 70% of Zn in the form of $ZnO_2$) in a muffle furnace in air at 570 K (120 min.). Nanopartcles with grain size of 10-50 nm can be clearly seen on the SEM image (Figure 1a). X-ray diffraction pattern of initial w-ZnO powder shows lines broadening due to nanocrystallinity (Figure 1d). No diffraction peaks from $ZnO_2$ or other crystalline impurities were detected.

Single-phase nanocrystalline bulk (pellet ~1 mm thickness and 4.5 mm in diameter) rs-ZnO has been synthesized from w-ZnO nanopowder (grain size of ~9 nm) at 7.7 GPa & 800 K and subsequent rapid quenching (Figure 1b & e). The details of high-pressure synthesis and characterization of the recovered samples are described earlier [4, 5, 7]. Nanocrystalline bulk w-ZnO obtained by reverse (from rs-ZnO to w-ZnO) phase transformation was used as reference sample (Figure 1c & f). This eliminates the potential influence of grain size and possible surface contribution to the heat capacity. Before and after the calorimetric measurement X-ray powder diffraction verified that the samples were single-phases without any impurity. As one can see on the upper right part of the photo (Figure 1b) between rs-ZnO nanograins exist intergrain boundaries that could contain water molecules absorbed on the w-ZnO surface before the high-pressure synthesis. As follows from Ref. [4], this water probably plays the role of "glue" and helps to stabilize metastable rock salt phase at ambient pressure. Elimination of these water molecules will inevitably lead to reverse transformation to wurtzite phase.

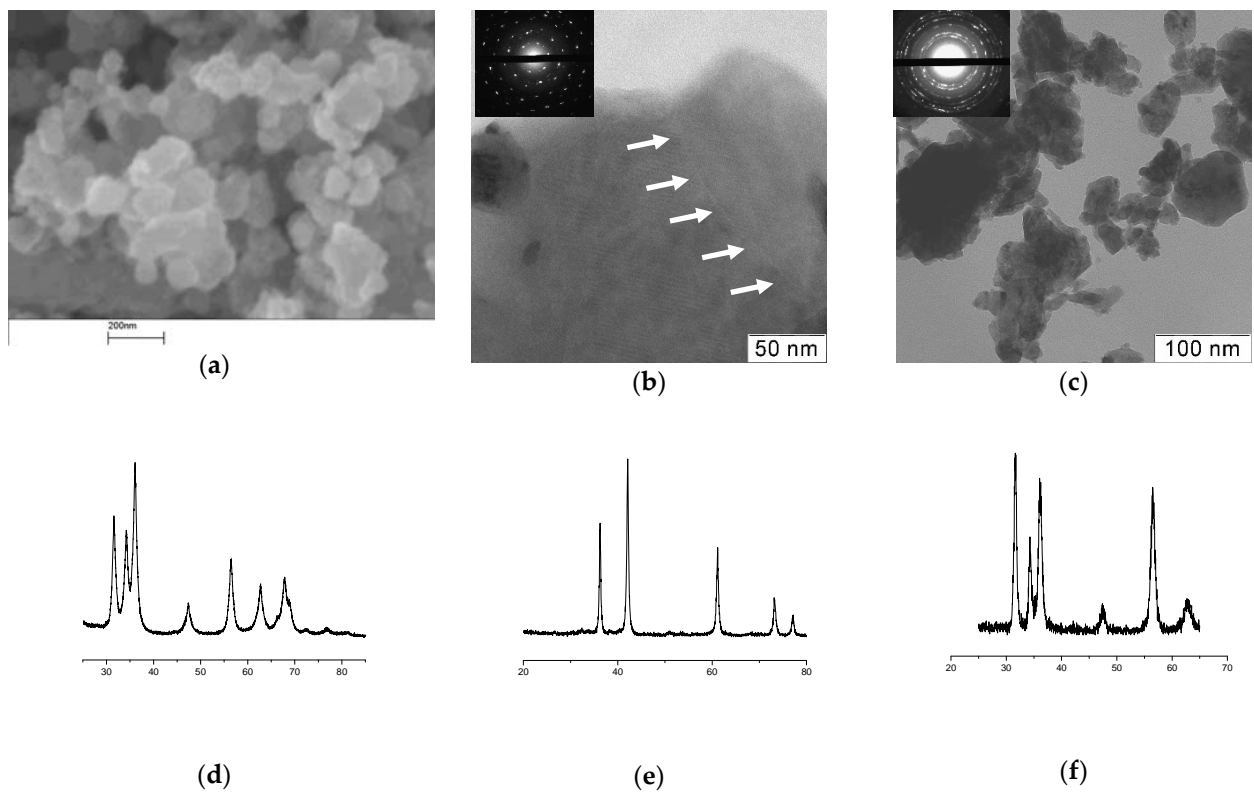

**Figure 1.** (**a**) SEM image of w-ZnO particles obtained by thermal decomposition of $ZnO_2$ that show morphology typical for nanopowders. (**b**) TEM image of rs-ZnO nanoparticles quenched from high pressure. Inset shows SAED pattern of rock salt structure. Arrows point to grain boundary. (**c**) TEM image of w-ZnO nanoparticles prepared by thermal decomposition of rs-ZnO. Inset shows SAED pattern of wurtzite structure. (**d**-**f**) X-ray diffraction patterns (CuK$\alpha$ radiation) of pristine w-ZnO nanopowder (**d**), synthesized rs-ZnO (**e**), and w-ZnO obtained by reverse phase transformation (**f**).

Heat capacity at constant helium pressure (35-60×10$^{-2}$ Pa) and in constant magnetic field (0.187 Oe) was measured at 40 different temperatures between 310 and 2 K, and two measurements were performed at each temperature. The data were collected using calibrated heater-thermometer platforms in Physical Property Measurement System apparatus by Quantum Design. The sintered pieces of rs-ZnO and w-ZnO were used for heat capacity measurements. One side of each piece was polished into a flat surface. Most of measurements were performed on two cubic-shaped samples with 4.90 mg (rs-ZnO) and 13.80 mg (w-ZnO) weight. The ZnO molar weight of 81.38 g/mole was used to molar $C_p$ calculation according to the 2011 IUPAC recommendation [8]. Recently it has been shown that PPMS can precisely measure the low-temperature $C_p$ for samples of milligram size [9]. Thus, the low-temperature $C_p$ measurements of high-pressure phases, which were almost impossible to achieve previously, now can be carried out using this technique. The accuracy of $C_p$ measurement in our experiments was estimated to be better than 1% between 4.2 K and 315 K, and 4% at temperatures below 4.2 K.

Calorimetric measurements were carried out using C80 CS Evolution Calvet calorimeter (SETARAM Instrumentation) equipped with 3D Tian–Calvet sensors. Calisto program package (ver. 1.086, AKTS AG) was used for running the experiment, data acquisition and initial data processing. Precise temperature calibration over the entire working range (293–540 K) for various modes was performed using the melting points of the standards (gallium, indium, tin); dynamic calibration was carried out by the Joule effect using E.J.3 unit and a dedicated calibration cell (S60/1434). It was additionally verified by measuring the melting heats of the standards.

The second series of calorimetric measurements was performed using a micro DSC-7 evo Calvet calorimeter (SETARAM Instrumentation) equipped with 3D Calvet-type sensors. Calisto program package (v 1.086, AKTS AG) was applied for running the experiment, data collection and initial data processing. Calorimeter calibration for heat production was done by the Joule effect using a E.J.3 calibration unit according to recommendations of SETARAM. Temperature calibration at various heating rates was performed with standard substances according to the IUPAC recommendations. Batch calorimetric vessels 1.0 mL made of Hastelloy C276 were used.

Stepwise heating mode was used for $C_p$ data collection as recommended for precise thermodynamic measurements [10]. Two consecutive runs with exactly the same heating modes (one – with the ZnO sample, another – without it – the base line) were performed. The durations of the linear heating step and isothermal step were selected taking into account the time constant of the calorimetric sensor along with the vessel. The first heating step was applied from 293.21 to 297.16 K ($C_p$ value at 295.125 K, isothermal step 20 min, then heating with scanning rate β = 0.20 K/min, heating step duration 20 min to reach equilibrium, then isothermal step 35 min), the second heating step was applied from 297.16 to 301.10 K ($C_p$ value at 299.13 K, scanning rate β = 0.20 K/min, heating step duration 20 min to reach equilibrium, then isothermal step 35 min). Subtraction of the corresponding heat effects was performed manually.

w-ZnO samples was obtained from the corresponding rs-ZnO samples via heating at controlled conditions in the course of DSC experiments and observing the heat generation at the phase transition using either DSC-131 evo scanning calorimeter or Labsys evo TG/DSC thermal analyzer (both by SETARAM Instrumentation). Linear heating (5 or 10 K/min) was applied.

DTG measurements employed for *in situ* probing the reverse rs-ZnO-to-w-ZnO transformation were also used for evaluation of the amount of absorbed water. Typically, the $H_2O$ loss in the range of 2-4 wt% depending on the sample was observed. These values can be considered as an underestimate, while the heat capacity contribution shows the water amount of ~5 wt% (this value should be taken with precaution since $C_p$ of absorbed water is higher than that of bulk water [11]). The correction to heat capacity has been made using the reference data for water in the 2–273 K and 273-310 K temperature ranges [12]:

$$^{H_2O}C_p \text{ (J mol}^{-1}\text{ K}^{-1}\text{)} = 18 \times (7.73 \times 10^{-3} \times T \times (1-\exp(-1.263 \times 10^{-3} \times T^2)) - 7.59 \times 10^{-3} + 2.509 \times 10^{-3} \times T - 1.472 \times 10^{-5} \times T^2$$
$$-1.617 \times 10^{-9} \times T^3 + 8.406 \times 10^{-11} \times T^4).$$

## 3. Results & Discussion

rs-ZnO samples available at ambient pressure are always nanostructured, and can survive only at quite moderate temperatures. Previous reports suggest possible impact of nanostructuring on ZnO heat capacity values. Our experiments, however, indicate that to the accuracy of the measurements, no difference between nanostructured w-ZnO (obtained by reverse transformation of rs-ZnO upon heating) and bulk w-ZnO is observed. At the same time, samples exposed to air may show higher heat capacity, if absorbed water correction is not made. Typically, 3-5 wt% of water was found as the equilibrium value for nano-rs-ZnO, and both nano-and micro-w-ZnO. Corrected values for rs-ZnO and w-ZnO are shown in Figure 2a. No distinction between nano- and micro- samples will be made in the following discussion.

The experimental values of the heat capacity for w-ZnO and rs-ZnO nanostructured samples (Figure 2b) were fitted to the adaptive pseudo-Debye model proposed by Holtzapfel (explicitly formulated in Ref. [13]) that gives analytical expression for $C_p$ as

$$c_p = 3R\tau^3 \frac{4C_0 + 3C_1\tau + 2C_2\tau^2 + C_3\tau^3}{\left(C_0 + C_1\tau + C_2\tau^2 + C_3\tau^3\right)^2} \left[1 + A\frac{\tau^4}{(a+\tau)^3}\right], \tag{1}$$

where $\tau = T/\theta_h$; $\theta_h$ is Debye temperature in the high-temperature region; $C_1$, $C_2$ and $A$ are parameters to be fitted; $C_3 = 1$; $a$ characterizes non-harmonicity; $R$ is gas constant. Parameter $C_0$ has been chosen as $C_0 = (5\ \theta_l^3)/(\pi^4\ \theta_h^3)$, with $\theta_l$ and $\theta_h$ Debye temperatures in the low- and high-temperature regions, respectively, in order to obtain these values directly as fitting parameters.

The fitting of the experimental data to Eq. 1 was performed using the simplex method using the MATLAB software. The uniqueness and stability of the solution (i.e. a set of mentioned above parameters determining theoretical curve) of inverse problem have been tested by multiple minimization procedures from various sets of starting parameters. To improve the solution quality and obtain the parameters allowing high-temperature extrapolation, we added the available data on $C_p$ of w-ZnO up to 1250 K (this allowed obtaining the reliable values of $\theta_h$, $A$ and $a$). It was possible to exclude the mutual compensation of fitting parameters only in the case of w-ZnO: the number and quality of the initial data combined with the $a$ value constrained to be positive and with taking into account the additional high-temperature data. In order to avoid occasional non-physical solutions for rs-ZnO, we used the w-ZnO parameters as the initial approximation. For both phases non-harmonicity parameter $a$ was found to be zero.

The fitting gave the following sets of parameters:

(1) for w-ZnO : $\theta_h$ = 256 K, $\theta_l$ = 132 K, $C_0$ = 0.0070, $C_1$ = 0.4490, $C_2$ = 0.9942, $A$ = 0.0208, $a$ = 0;

and

(2) for rs-ZnO : $\theta_h$ = 226 K, $\theta_l$ = 335 K, $C_0$ = 0.1663, $C_1$ = 0.3864, $C_2$ = 0.9755, $A$ = 0.0055, $a$ = 0.

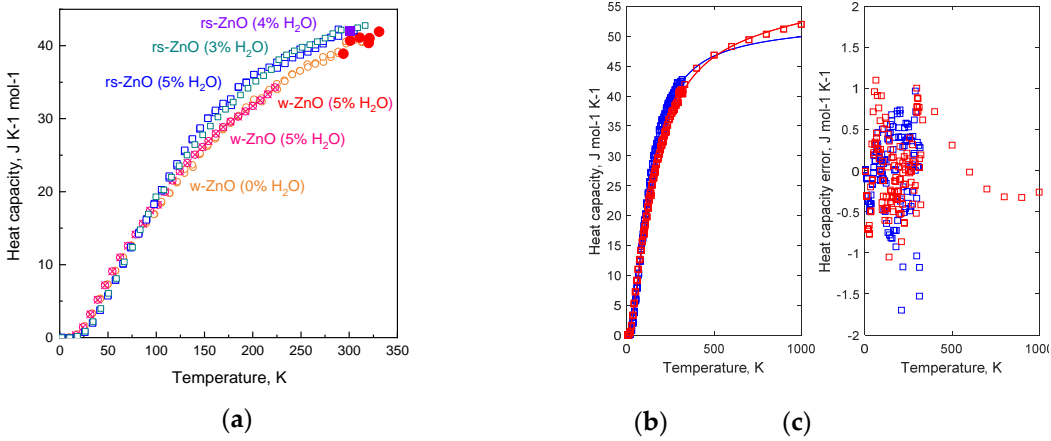

**Figure 2. (a)** Experimental heat capacities of nanostructured rs-ZnO (squares) and w-ZnO (circles). Open symbols correspond to PPMS measurements, while solid symbols – to SETARAM calorimeter measurements. **(b)** Heat capacity data fitted to Holtzapfel equation (1). High-temperature extrapolation for rs-ZnO has been made using regularized fitting procedure described in the text. Red and blue solid lines represent the calculated $C_p$ curves for w-ZnO and rs-ZnO, respectively. Solid squares of the same color show experimental data. The size of squares corresponds to the error estimate. **(c)** Discrepancy between experimental and calculated values of heat capacities. Most of the data points are in within ±0.5 J mol$^{-1}$ K$^{-1}$.

One can see that the high-temperature limits of Debye temperature $\theta_D$ are close, while the low-temperature values are in agreement with the general rule that the higher density corresponds to higher Debye temperature. At the same time, our $\theta$-values are noticeably lower than the previously reported Debye temperatures (~400 K) for w-ZnO [14]. Such discrepancy can be explained, from one side, by nanostructuring of our samples, and, from the other side, by the experimental errors that allow treating the $\theta_l$ values as fitting parameters. Only rigorous study of fully water-free samples could reveal whether nanostructuring impacts Debye temperature or not.

In the case of the nanostructured w-ZnO obtained by reverse transformation from rs-ZnO at 523 K, the heat capacity is quasi-indistinguishable from the large-grain samples. This is indicative of essentially zero surface entropy of our nano-w-ZnO, similar to previous observations for nano-w-ZnO obtained by chemical decomposition of zinc nitrate at 573 K [11]. Such closeness can be attributed to the thermal cure of the surface.

Figure 3a shows the formation enthalpy $\Delta H_f$ calculated using the fitted values of heat capacities and experimental value of $\Delta H_f$ (298.15 K) = 11.7±0.3 kJ mol$^{-1}$ [6]. One can observe only weak temperature dependence of $\Delta H_f$. The minimum and maximum on the curve are due to the double intersection of heat capacity curves of w-ZnO and rs-ZnO nanostructured phases. These values differ by a factor of two from the results of indirect measurements at high temperatures (e.g. by EFM measurements [15]), which can be explained by the 1100-1300 K experimental range that is out of rs-ZnO stability region, as well as by the nanocrystallinity of our samples. It is also interesting to compare $\Delta H_f$ values with the *ab initio* predictions (at 0 K) that show a large dispersion (from 15 to 30 kJ mol$^{-1}$) of calculated values, e.g. 21.230 and 28.950 kJ mol$^{-1}$ (by LDA and GGA, respectively) in Ref. [16] and 15.247 and 22.871 kJ mol$^{-1}$ (by LDA and GGA, respectively) in Ref. [17]. The variation of the calculated values by a factor of two does not allow making any reliable evaluation of thermodynamic stabilities of bulk phases at low temperatures. At the same time, the *ab initio* simulations of structural transition in ZnO nanowires at high pressures (calculations using the SIESTA code) indicated that passage from bulk crystal to nanograins reduces $\Delta H_f$ (0 K) from 23 kJ mol$^{-1}$ down to 10 kJ mol$^{-1}$ [18]. This result is consistent with the data in Ref. [6] and imply the difference between bulk- and nano-phase diagrams of ZnO.

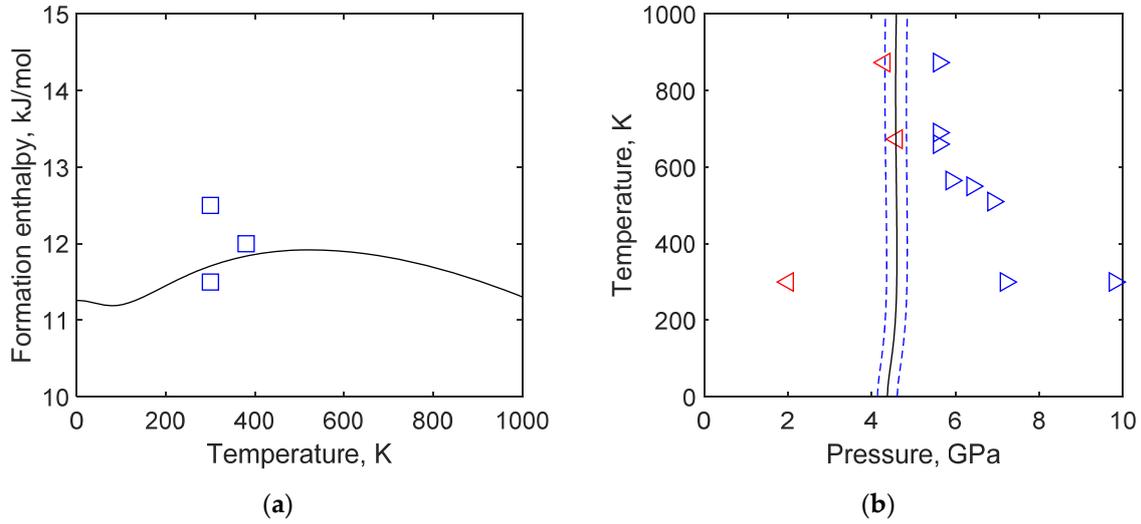

**Figure 3.** Calculated $\Delta H_f(T)$ at 0.1 MPa and $p$-$T$ phase diagram (i.e. $\Delta G_{rs/w}(p,T) = 0$) vs experimental and *ab initio* data. (**a**) Black solid line is $\Delta H_f$ as a function of temperature extrapolated to 1000 K using regularized procedure described in the text; blue open squares are experimental data from Ref. [6]; (**b**) Nano-phase diagram of ZnO representing equilibria between nanostructured phases. Isopotential curve corresponds to $\Delta G_{rs/w}(p,T) = 0$ (equilibrium between w-ZnO and rs-ZnO; black solid line) with the estimated error of ±0.3 kJ mol$^{-1}$ for $\Delta H_f$ [6] (the dashed blue lines). Symbols represent experimental $p,T$-points on direct (right-oriented red triangles) and inverse (left-oriented blue triangles) transformations observed by *in situ* X-ray diffraction [7,19].

Nano-phase diagram of zinc oxide has been calculated using the fitted experimental heat capacity values, directly-measured $\Delta H_f$ [6] and thermoelastic $p$-$V$-$T$ data on rs-ZnO [5,19] and w-ZnO [20]. The experimental data on direct and inverse transformations in ZnO at high (above 500 K) temperatures are in satisfactory agreement [7,19]. At lower temperatures, where the transformation hysteresis is pronounced, the application of the estimated $p_{eq}$ for the equilibrium pressure as

$$p_{eq} \approx 0.5 \times (p_{direct} + p_{inverse}) \quad (2)$$

results in the $p_{eq}$ value of 5±3 GPa at 300 K [19,21]. The uncertainties have an order of magnitude of the yield stress accumulated during the transformation [22]. Here we should note that the use of the hypothesis that $p_{eq} \approx p_{direct}$ is methodologically incorrect. In fact, both direct and inverse transformations were observed experimentally at finite pressure, which is in agreement with Landau character of 2$^{nd}$-order transformation driven by the lattice strain as order parameter. Such model requires the onset pressure of ~9.4 GPa [22] above the equilibrium corresponding to the formal requirement of $\Delta G_{rs/w}(p,T) = 0$.

## 5. Conclusions

The low-temperature (2–310 K) isobaric heat capacities of rock salt ZnO, metastable high-pressure phase, and wurtzite ZnO have been experimentally studied using the thermal relaxation PPMS calorimetry. Measurements of low-temperature heat capacities of nanostructured ZnO polymorphs and analysis of the equation-of-state data allowed us to resolve the ambiguities of the equilibrium $p$-$T$ phase diagram of ZnO. A set of proposed thermo-physical data is well consistent with the observed direct and inverse phase transformations in nanostructured ZnO up to 1000 K. In this temperature range the equilibrium pressure is close to 4.6 GPa, which is lower than the *ab initio* prediction for bulk ZnO, but in a good agreement with simulations made for nanoparticles.

**Author Contributions:** Conceptualization, V.L.S. and A.C.; methodology, A.C. and V.L.S.; investigation, K.V.K., P.S.S., A.N.B., F.Y.Sh. and V.L.S.; data curation, K.V.K. A.C. and V.L.S.; writing—original draft preparation, A.C.; writing—review and editing, V.L.S.; visualization, A.C.; supervision, V.L.S.; funding acquisition, V.L.S. All authors have read and agreed to the published version of the manuscript.

**Funding:** This research was funded in part by the Russian Foundation for Basic Research, grant number 11-03-01124.

**Acknowledgments:** P.S.S. is thankful to the "Science and Engineering for Advanced Materials and devices" (SEAM) Laboratory of Excellence for financial support. A.N.B. is grateful to the Université Sorbonne Paris Cité for financial support.

**Conflicts of Interest:** The authors declare no conflict of interest.